\documentclass[article,balancelastpage,twocolumn,prl]{revtex4}%
\usepackage{makeidx}
\usepackage{amssymb}
\usepackage{dcolumn}
\usepackage{graphicx}
\usepackage{acronym}
\usepackage{amsmath}
\usepackage{amsfonts}%
\ifx\pdfoutput\relax\let\pdfoutput=\undefined\fi
\newcount\msipdfoutput
\ifx\pdfoutput\undefined\else
\ifcase\pdfoutput\else
\ifx\paperwidth\undefined\else
\ifdim\paperheight=0pt\relax\else\pdfpageheight\paperheight\fi
\ifdim\paperwidth=0pt\relax\else\pdfpagewidth\paperwidth\fi
\begin{document}
	
\title{Is our space an object of point rotation?}
\author{B. V. Gisin }

\affiliation{E-mail: borisg2011@bezeqint.net}
\date{\today }
	
\begin{abstract}
The interval of a flat Minkowski space is invariant with respect to the previously found three-dimensional transformation for point rotating coordinate systems. The assumption that our space is an object of point rotation at a frequency initiates the appearance of a new term in the Dirac equation. The upper boundary for this frequency is determined from shifts of the degenerate levels of hydrogen atom. The boundary is about 2.9 kHz.
		
Keywords: Point rotating frames, Dirac's equation, Lamb shift.  
\end{abstract}.
	
\maketitle
	
\section{Introduction}

An example of a pointwise (below the point) rotation can be found in a circularly polarized electromagnetic wave. Its distinctive feature is the existence of a rotation axis at each point. The observer can not be placed in the frame with an infinite number of  rotating axes. He can operate with such frames only mentally.

Point rotation is fundamentally different from mechanical. It is important to emphasize, because here we are considering a touchstone of quantum mechanics - the motion of an electron in a hydrogen atom. 

In accordance with the mechanical description, the angle of rotation in an elliptical orbit depends on the radius, whereas in the quantum-mechanical states of a hydrogen atom this angle and radius vary independently. Therefore, below in the transformation for point rotating coordinate systems (rotating frames), the independence criterion should be used for the angle and radius.

The problem of transition from one point rotating frame to another arises, in particular, in optics, describing the action of an electrooptical single-sideband modulator. In such a modulator, a rotation of the optical indicatrix (index ellipsoid) is used under the influence of a rotating electric field.
This modulator shifts the optical frequency without harmonics \cite{BB}, \cite{CS} and can be used to test the superposition of angular velocities \cite{BG}. 

Unfortunately, at present the accuracy of optical measurements is insufficient to detect a difference from simple frequency addition.

A similar problem exists in the classical description of magnetic resonance in the framework of Pauli's equation. In both cases, a simple addition of frequencies is assumed, and time in the rotating frame is the same as in the stationary frame.

It is well known that a three-dimensional transformation of the spatial Cartesian coordinates can be obtained by sequentially rotating three pairs of coordinates. This transformation depends on three parameters (angles).

In \cite{pr} 3D transformation for rotating frames was obtained by sequentially rotating three pairs of linear elements (differentials) of cylindrical coordinates, an angle, a distance along the rotation axis and time.

For these rotations two additional conditions were used in \cite{pr}. The first is constancy of the speed of light along the axis of rotation, the second is absence of singularities for all allowable frequencies in the forward and  backward transformations.

As a result, the 3D transformation depends on the frequency and a constant. The independence criterion of the angle on radius is satisfied in this transformation. Its surprising property is that the interval of a flat Minkowski space is invariant under the transformation. This means that we must not exclude the possibility that our space is an object of point rotation, i.e., a relic frequency (this term is used for certainty) is not necessarily zero. Moreover, the wave equation is also invariant with respect to this transformation.

Unlike the interval, Dirac's equation is not invariant under the 3D transformation. The equation after this transformation acquires a new term, which violates its symmetry. 

Another surprise is connected with the consideration of magnetic resonance within the framework of Dirac's equation in a rotating electromagnetic field. The magnetic moment begins to rotate together with the frame. In this connection, the magnetic resonance condition should be considered as the solutions  stability in this rotating frame.
In \cite{pr} it was shown that this condition in the nonrelativistic approximation exactly coincides with the classical condition when the $ g $-factor coincides with the constant in the 3D transformation.

In the paper, we assume that space is an object of the point rotation at a relic frequency. We study the influence of new term in modified Dirac's equation on the shift of hydrogen degenerate levels.

We show that the structure of this shift differs radically from the Lamb shift, which is well described by quantum field theory in accordance with the experiments.

The purpose of this article is rather utilitarian. Namely, the definition of the upper boundary of the relic frequency according to experimental or theoretical data.

\section{3D Transformation for rotating frames}

Assuming that 'our space' is an object of point rotation, we must also postulate that there is a space in which the relic frequency is zero, and all physical laws have the usual form. Then, if the physical law is not invariant with respect to the 3D transformation (as in the Dirac equation), an additional term appears that depends on the relic frequency.

In connection with importance of the issue we repeat here the derivation of the three-dimensional transformation obtained by successive rotations of three pairs of differentials of cylindrical coordinates $rd\varphi, dz, cdt$ \cite{pr}.

The first rotation of the pair ($rd\varphi, cdt$) is as follows
\begin{align}
& rd\varphi^{\prime} =rd\varphi\cosh\Phi-cdt\sinh\Phi,\label{Tp1} \\
& \text{ }c dt^{\prime} =-rd\varphi\sinh\Phi+c dt\cosh\Phi,\label{Tt1} 
\end{align}
where $r\omega$ is the line velocity on the circle of radius $r$, $\omega$ is the frequency.

In fact (\ref{Tp1}), (\ref{Tt1}) is the Lorentz transformation.
For small velocity $|\omega r|\ll c$ this turns into a
non-relativistic rotation: $d\varphi'=d\varphi-\omega dt, $ $dt'=dt$. Therefore we assume that
\begin{equation}
\tanh\Phi =\frac{r\omega}{c}, \label{tan}%
\end{equation}
where $\omega$ is the relic frequency.

 We use here the term 'rotation' for the all three cases.

Obviously, the rotations (\ref{Tp1}), (\ref{Tt1}) together with next two describes transition into a rotating non-inertial frame with centrifugal forces that depend on $r$. Therefore, we assume that all angles of rotation
$\Phi, \; \Phi_1, \; \Phi_2$ depend on $r$. This in a sense is equivalent to the action of centrifugal forces. 

The second pair of differentials is $(dz, rd\varphi')$. The rotation of this pair is
\begin{align}
dz' &  =dz\cos\Phi_{1}-rd\varphi^{\prime}\sin\Phi_{1},\label{Tp2}\\
\text{ \ }rd\tilde{\varphi}  &  =dz\sin\Phi_{1}+rd\varphi^{\prime}\cos\Phi
_{1}, \label{Tt2}%
\end{align}
This rotation corrects the direction of $z'$ axis.

The third pair is ($dz^{\prime},dt^{\prime}$). This rotation corrects momentum along the new $\tilde{z}$ axis
\begin{align}
d\tilde{z}  &  = dz^{\prime}\cosh\Phi_{2}-c dt^{\prime}\sinh\Phi_{2}%
,\label{Tp3}\\
c d\tilde{t}  &  =-dz^{\prime}\sinh\Phi_{2}+c dt^{\prime}\cos\Phi_{2},
\label{Tt3}%
\end{align}

The denominator 
of $\sinh\Phi$ and $\cosh\Phi$ in (\ref{Tp1}),(\ref{Tt1}) is $\sqrt{1-\omega^{2}r^{2}/c^{2}}$.
It means that the maximal value of $r$ is bounded:
\begin{equation} 
r\leq\ \frac{\lambda}{2\pi}, \label{rM}
\end{equation}
where $\lambda=2c\pi/\omega$ is the wavelength corresponding to the circular frequency $\omega$.

The angle $\Phi$ in 3D transformation is specified by (\ref{tan}). $\Phi_1$ is defined by the condition of speed of light constancy 
\begin{align}
\frac{d\tilde{z}}{d\tilde{t}}=c \quad \mathrm{if} \quad \frac{dz}{dt}=c. \label{cc}
\end{align}
From this condition we get the connection between $\Phi_{1}$ and $\Phi$%
\begin{equation}
\sin\Phi_{1}=\tanh\Phi=\omega r/c, \label{sc}%
\end{equation}

Because of the condition (\ref {cc}), the 3D transformation remains unchanged if we swap places of the first two rotations.

The 3D transformation has singularities at $|\omega r|=c.$ The singularity vanishes and the transformation is simplified, if 
$\Phi_{2}$ is defined by only one condition%
\begin{equation}
\Phi_{2,r}=\Phi_{,r}\sin\Phi_{1},\text{ }\exp\Phi_{2}=\frac{\varkappa}%
{\sqrt{1-\Omega^{2}r^{2}/c^{2}}}. \label{f2}%
\end{equation}

3D transformation, as the coordinate dependence
$(rd\tilde{\varphi}, d\tilde{z}, cd\tilde{t})$ on $(rd\varphi, dz, cdt)$ can be obtained after excluding the primed coordinates from above rotations.

After the excluding, 3D transformation may be written as follows
\begin{align}
d\tilde{\varphi}  &  =d\varphi+\frac{\omega}{c}z-\omega dt,\label{tphi}\\
d\tilde{z}  &  =-\frac{\omega r^{2}}{c\varkappa}d\varphi+\varkappa_{22}%
dz+\varkappa_{23}cdt,\label{tz}\\
cd\tilde{t}  &  =-\frac{\omega r^{2}}{c\varkappa}d\varphi+\varkappa_{32}%
dz+\varkappa_{33}cdt. \label{tt}%
\end{align}

Obviously, the criterion of independence $\tilde\varphi $ on $ r $ is satisfied.

The reverse transformation is
\begin{align}
d\varphi  &  = d\tilde{\varphi}-\frac{\omega}{c\varkappa}d\tilde{z}+\frac{\omega}{\varkappa} d\tilde{t},\label{rphi}\\
dz  &  = \frac{\omega r^2}{c}d\tilde{\varphi}+\varkappa_{22}%
d\tilde{z}-\varkappa_{32}cd\tilde{t},\label{rtz}\\
cdt  &  = \frac{\omega r^2}{c}d\tilde{\varphi}-\varkappa_{23}%
d\tilde{z}+\varkappa_{33}cd\tilde{t}. \label{rt}%
\end{align}
$\varkappa$ is the dimensionless constant. It is shown in \cite{pr} that $\varkappa$ 
is nothing but the $g$-factor in the expression of anomalous magnetic moment. $\varkappa_{kl}$ is defined as follows
\begin{align}
\varkappa_{22}  &  =\frac{1}{2\varkappa}(1+\varkappa^{2}-\frac{1}{c^{2}}%
\omega^{2}r^{2}),\label{C22}\\
\varkappa_{23}  &  =\frac{1}{2\varkappa}(1-\varkappa^{2}+\frac{1}{c^{2}}%
\omega^{2}r^{2}),\label{C23}\\
\varkappa_{32}  &  =\frac{1}{2\varkappa}(1-\varkappa^{2}-\frac{1}{c^{2}}%
\omega^{2}r^{2}),\text{ }\label{C32}\\
\varkappa_{33}  &  =\frac{1}{2\varkappa}(1+\varkappa^{2}+\frac{1}{c^{2}}%
\omega^{2}r^{2}). \label{C33}%
\end{align}

The determinant of 3D transformation, as well as its reversal, is 1. 

It is easy to verify that 4D interval as well as 4D differential form
\begin{align}
& ds^2=dr^2+r^{2}d\varphi^{2}+dz^{2}-c^{2}dt^{2}, \label{4i} \\
& \frac{\partial^2}{\partial{r^2}}+\frac{\partial}{r\partial r} +\frac{\partial^2}{r^{2}\partial\varphi^2}+\frac{\partial^2}{\partial z^2}-\frac{\partial^2}{c^{2}\partial{t^2}} \label{we}
\end{align} 
are invariants.  

We consider below levels of the hydrogen atom. Therefore we should define the electric potential. Since the operator (\ref{we}) is invariant, the form of the Coulomb potential, as the static spherical symmetric solution of the wave equation, does not change, in both the resting and rotating frame. Only coordinates changed
\begin{align}
V=-\frac{e^2}{\rho}, \;\; \rho^2=\tilde{x}^2+\tilde{y}^2+\tilde{z}^2, \nonumber 
\end{align}
where spherical coordinate determined as usually $\tilde{x}=\rho\sin\theta\cos\tilde{\varphi}, \;\; \tilde{y}=\rho\sin\theta\sin\tilde{\varphi}, \;\; \tilde{z}=\rho\cos\theta.$

Below we drops the tilde over $\varphi$
 
\section{The Dirac equation}

In this section we use the normalized units 
\begin{equation} 
\rho=\frac{r}{\lambdabar}, \;\; \alpha=\frac{e^2}{\hbar c}, \;\; \lambdabar=\frac{\hbar}{mc}, \;\; \varepsilon=\frac{E}{mc^2}, \;\; \tau=\frac{\hbar\omega}{2\varkappa mc^2} \nonumber
\end{equation}
where $\tau$ is the small dimensionless parameter, defining the quantity of the new term in Dirac's equation, $\epsilon$ is energy.

Dirac's equation in a rotating electromagnetic field was found in \cite{pr}. We can use this result, equating this field to zero and its frequency to a relic frequency. The Dirac equation in the static case takes the form
\begin{align}
& [D(\tilde{x},\tilde{y},\tilde{z})-\varepsilon+\beta+\tau\gamma_a]\tilde{\Psi}=0, \label{eqn}
\end{align}
where the operators $D(\tilde{x},\tilde{y},\tilde{z})$ in the spheric coordinates and $\gamma_a$ are as follows
\begin{align}
& D=-i\left(\alpha_1\frac{\partial}{\partial r} +\alpha_1\frac{1}{2r}+%
\alpha_2\frac{\partial}{r\partial\tilde{\varphi}}+%
\alpha_3\frac{\partial}{\partial\tilde{z}}\right), \label{L} \\
& \gamma_a=-i(1-\alpha_3)\alpha_1\alpha_2=\begin{pmatrix} 1 & 0 & -1 & 0 \\ 0 & -1 & 0 & -1 \\ -1 & 0 & 1 & 0 \\ 0 & -1 & 0 & -1  \end{pmatrix}.  \label{mm}
\end{align}

\section{Solutions}

We use well known exact solutions of the Dirac's equation (\ref{eqn}) with the Coulomb potential, described in different textbooks and monographes \cite{AxB}, as zero approximation at $\omega=0$. Further we find approximate solutions of the modified Dirac's equation (\ref{eqn}) with the additional term in the framework of perturbation theory.

The matrix $\gamma_a$ mixes components of the wave function $\tilde{\Psi}$, therefore the parameter $\tau$ must be very small $|\tau| \ll 1-\epsilon \ll 1$  

\subsection{Exact solution}

As it is known the Hamiltonian of Dirac's equation at $\omega=0$ commutes with the components of total angular moment $J_k$. If $\omega \ne 0$ then the Hamiltonian commutes only with  $J_3$  
\begin{align}
J_3=-i\frac{\partial}{\partial\varphi}+\frac{1}{2}\sigma'_3, \quad \frac{1}{\hbar}J_3\tilde{\Psi}=j_3\tilde{\Psi}, 
\end{align}
where $\sigma'_1=-i\alpha_2\alpha_3, \; \sigma'_2=-i\alpha_3\alpha_1, \; \sigma'_3=-i\alpha_1\alpha_2$, $j_3$ is the half-integer eigenvalue.

This allows us to find dependence of the wave function on the spherical angle $\varphi$.
\begin{align}
\tilde{\Psi}=\exp[(ij_3-\frac{1}{2}\alpha_1\alpha_2)\varphi]\psi(\rho,\theta). \label{Psi} 
\end{align}
where $\mu=j_3-1/2$ is an integer.

Moreover the Hamiltonian commutes with the operator $K$
\begin{align}
K=\beta(\sigma'_k \frac{L_k}{\hbar}+1), \;\; \mathbf{L}=-\mathbf{[xp]}, \label{ek}
\end{align}
where eigenvalues of this operator are $k=\pm 1,\pm 2, \pm 3,$ ...

Excluding with help of (\ref{Psi}) the dependence on $\varphi$ from the equation
\begin{align}
k\tilde{\Psi}=\beta(\sigma'_k \frac{L_k}{\hbar}+1)\tilde{\Psi}, 
\end{align}
we can find that the dependence of all components $\psi_k$ on $\theta$ is defined by the Legendre polynomials.
  
The sign change of $k$ corresponds to the transformation $\psi'=-i\alpha_1\alpha_2\alpha_3\psi$.

The two exact solutions are
\begin{align}
k=l>0, \; \Psi_{+}=\begin{pmatrix} & F_{l-1}(l+\mu)P^{\mu}_{l-1}\exp{i\mu\varphi} \\ &  F_{l-1}P^{\mu+1}_{l-1}\exp{i(\mu+1)\varphi} \\ & -iG_{l}(l-\mu)P^{\mu}_{l}\exp{i\mu\varphi} \\ &  iG_{l}P^{\mu+1}_{l}\exp{i(\mu+1)\varphi} \\ \end{pmatrix}, \label{Cm1} \\
k=-l<0, \; \Psi_{-}=\begin{pmatrix} & -iG_{l}(l-\mu)P^{\mu}_{l}\exp{i\mu\varphi} \\ &  iG_{l}P^{\mu+1}_{l}\exp{i(\mu+1)\varphi} \\ & 
F_{l-1}(l+\mu)P^{\mu}_{l-1}\exp{i\mu\varphi} \\ & 
F_{l-1}P^{\mu+1}_{l-1}\exp{i(\mu+1)\varphi} \\ \end{pmatrix}. \label{Cm2}
\end{align}
where $F(\rho)$, $G(\rho)$ are some functions, $P^{0}_0=1, \; \mu\leq\l$. 

The angular part of the wave functions can be normalized with help of the relation
\begin{align}
& [(l+\mu)P^{\mu}_{l-1}]^2+[P^{\mu+1}_{l-1}]^2= 
2\frac{(l+\mu)!}{(l-\mu-1)!},  \nonumber \\
& [(l-\mu)P^{\mu}_{l}]^2+[P^{\mu+1}_{l}]^2= 
2\frac{(l+\mu)!}{(l-\mu-1)!},  \nonumber
\end{align}
and the complete wave function with help of the normalization integral over all the volume of the field
\[ \int\Psi^{*}_{-}\Psi_{-}dv=\int\Psi^{*}_{+}\Psi_{+}dv=1\]

Substituting $\Psi_{+}$ or $\Psi_{-}$ in the Dirac's equation, we obtain two equations for $F(\rho)$ nd $G(\rho)$. 

The requirement of square integrable solutions of the equations gives the third quantum number $n_r$, in additional to $j_3, k$, and the expression for energy
\begin{align} 
& \epsilon=\sqrt{1+(\frac{\alpha}{\gamma+n_r})^2}, \label{E} 
\end{align}
where $\gamma=\sqrt{k^2-\alpha^2}$.

\subsection{Shift of levels}

We now consider the modified Dirac equation with a small additional 'potential' $V_{\tau}=\gamma_\tau$, associated with a small change in energy. This term shifts the degenerate levels. 

The eigenfunctions  $\Psi_{+}, \; \Psi_{-}$ are orthogonal, they obey the equation
\begin{align} 
& \epsilon\Psi_{\pm}=H\Psi_{\pm}, \quad \Psi_{-}=-i\alpha_1\alpha_2\alpha_3\Psi_{+}, \nonumber 
\end{align} 
Accordingly to the perturbation theory we are looking for solutions of the equation
\begin{align} 
(\epsilon+\delta\epsilon)\Psi=(H+V_{\tau})\Psi \label{gEP}
\end{align}
in the form $[C_{+}\Psi_{+}+C_{-}\Psi_{-}+\sum c_k\Psi_k]$, where terms with $\Psi_{+},\; \Psi_{-}$  are written singly, $\delta\epsilon$, is the small addition to the  energy level, $c_k \ll C_{\pm}$. 

Taken into account the wave functions orthogonality,
multiply Eq. (\ref{gEP}) by $\Psi^{*}_{+}$ and $\Psi^{*}_{-}$ and integrating over all volume of the field we find two solutions of the secular equation in first approximation
\begin{align}
\delta\epsilon_{\pm}=\int\Psi^{*}_{+}\sigma'_3\Psi_{+}dv \pm\tau \label{se1}.
\end{align}
Finally the shift of levels is $\delta\epsilon_{+}-\delta\epsilon_{-}= 2\tau$ or in the non-normalized units
\begin{align}
\delta E_{+}-\delta E_{-}=\frac{\hbar\omega}{\varkappa}.
\end{align}

Using uncertainty of Lamb's shift measurements $L_{2s-2p}$ = 1057.8446(29) MHz \cite{L}, we can estimate the upper boundary of the relic frequency. 

Taken into account that  $\omega=2\pi \nu$ is the angular frequency and for electron $\varkappa \approx 1$, we obtain this boundary 
\begin{align}
\nu<2.9 \;\text{kHz}
\end{align} 

At a very small value of $\nu$, the influence of the relic frequency should be sought in cosmic phenomena

\section{Discussion}

The above invariance returns us to the Einstein idea of the non-symmetric metric tensor. Its antisymmetric part is associated with the electromagnetic field.

In \cite{prb}, for simplicity, the Minkowski space with electromagnetic field has been considered as the first approximation in the framework of general relativity. The gravitational field is zero in this first approximation.  

An algorithm for constructing the tensor energy momentum was presented. In this construction, the current terms are inserted into the tensor, while the Einstein tensor consists of the field terms. This construction executed so that the Einstein equation for the antisymmetric part of metric coincides with the Maxwell equation and all that can be done covariantly in each order the successive approximations.

The gravitation comes into existence in the second approximation, when the quadratic current terms arise in the tensor momentum. Shape of this term coincides with that of the gravitation term
\begin{align}
(C_m\mu+C_e\rho^{2})\frac{dx_\sigma}{dt}\frac{dx_\nu}{dt}
\end{align}
where $\mu$ and $\rho$ are the mass and charge density, $C_m, \; C_e$ are constants. It means that the charge density squared is proportional to the mass density.

Surprisingly, this relation can be found in the Bethe-Weizs{\'a}cker semi-empirical mass formula for atomic nucleus. However, the sign of $C_e$ is negative in this formula. 

This relation, if it is corresponding to reality, would have far-reaching consequences. In particular, the increase in pressure, so that the charge density becomes more nuclear, can be the cause of the Big Bang.
 
At this time, apparently, the relic frequency was great.

\end{document}